\documentclass[letterpaper,10pt]{article}
\usepackage{osameet2}

\pdfoutput=1

\usepackage{amsmath,amssymb}

\usepackage{graphicx}
\usepackage{tikz}
\usepackage{pgfplots}
\usepgfplotslibrary{colorbrewer}

\pgfplotsset{compat=1.14}

\usepackage{comment}

\begin{document}

\title{74.38~Tb/s Transmission Over 6300~km Single Mode Fiber with Hybrid EDFA/Raman Amplifiers}

\author{
    M.~Ionescu\textsuperscript{1,*},
    D.~Lavery\textsuperscript{2},
    A.~Edwards\textsuperscript{1},
    E.~Sillekens\textsuperscript{2},
    L.~Galdino\textsuperscript{2}, 
    D.~Semrau\textsuperscript{2},
    R.I.~Killey\textsuperscript{2},
    W.~Pelouch\textsuperscript{3},
    S.~Barnes\textsuperscript{1},
    and
    P.~Bayvel\textsuperscript{2}
}
\address{
    \textsuperscript{1} Xtera, Bates House, Church Road, Romford, RM3 0SD, UK.\\
    \textsuperscript{2} Optical Networks Group, University College London, Torrington Place, London,  WC1E 7JE, UK.\\
   \textsuperscript{3} Xtera, 500 West Bethany Drive, Allen, Texas 75013, USA.
    }
\email{*maria.ionescu@xtera.com}

\begin{abstract}
Transmission of 306$\times{}$35~GBd, dual polarization, 64-\textit{ary} geometrically shaped channels over 90$\times{}$70~km of SMF was demonstrated, achieving a net throughput of 74.38~Tb/s. A combination of hybrid fiber spans and EDFA/Raman amplifiers enabled a continuous gain bandwidth of 10.8~THz.
\end{abstract}


\section{Introduction}
Research in long-haul optical fiber communication systems is currently being driven by two complementary strategies: bandwidth expansion and spectral efficiency enhancement. 
For example, recent record capacity demonstrations have implemented extended bandwidth long-haul transmission using both C- and L-band EDFAs \cite{ghazisaeidi,cai}, effectively doubling the transmission bandwidth, albeit with a spectral guard band between the transmission windows. Semiconductor optical amplifiers (SOA) have also been considered for continuous-band gain in transmission systems with bandwidths up to 100~nm\cite{renaudier}. However, SOAs typically exhibit a higher noise figure than EDFAs, and so have not yet been demonstrated in cascade for long-haul systems.

In terms of improving the spectral efficiency, a common technique for long-haul systems is coded modulation. The simultaneous use of multiple code rates for forward error correction (FEC) has been shown to increase transmission throughput, by maximising the information rate on a per-channel basis. Recently, probabilistically (PS) \cite{ghazisaeidi} and geometrically shaped (GS) \cite{zhang} modulation formats have also been proposed to increase the achievable information rate (AIR) for a given signal-to-noise ratio (SNR). However, signals shaped to maximize AIR in an additive white Gaussian noise channel (AWGN) typically lose the features exploited by receiver-side digital signal processing (DSP) algorithms (e.g., carrier phase estimation and equalization) and therefore require a training overhead, which in turn reduces the gains from shaping.

In this paper, we used hybrid EDFA/Raman amplification to increase signal bandwidth, while enabling an increased span length and SNR versus the equivalent system based purely on lumped amplification. To further increase the AIR, we designed and implemented a new GS constellation which maximized the AIR, taking into account both the physical transmission channel and the DSP. We achieved a rate of 74.38~Tb/s, a record net throughput for single-mode fiber over a transoceanic distance of 6300 km.

\section{Amplifier Design}
The hybrid EDFA with distributed Raman amplifiers used in this experiment provided continuous gain from 1525~nm to 1616~nm.
The wide bandwidth was achieved by using two counter-propagating pumps at 1427~nm and 1495~nm with output powers of 300~mW and 310~mW into the transmission fiber, delivering a total signal power of 19.5~dBm to the EDFA stage. The single EDFA stage boosted the signal to a total output power of 22~dBm, and included a 91 nm gain flattening filter (GFF) to equalize the gain across the entire amplifier's bandwidth. The distributed Raman amplification stage reduced the effective noise figure to an average of 1.4 dB, thus enabling extra long fiber spans of 70~km, allowing an additional ~3dB span loss (or 32.6$\%$ span length extension) compared to previous experiments \cite{cai}.
The noise figure tilt was -5.7~dB but a -2~dB signal spectral tilt was chosen in order to take into account the wavelength-dependent nonlinearities such that the nonlinear penalty was approximately constant across the entire band.

%
%
%
%

\section{Pragmatic Geometric Shaping} 
The geometrically shaped modulation format used in this work was designed by initially using a gradient descent algorithm with a cost function seeking to maximize generalized mutual information (GMI) at an SNR of 12~dB. The resulting constellation produced an equivalent GMI to the formats designed in \cite{TUE_ECOC}. However, to maximize the GMI in a practical transmission system, the design algorithm was further constrained to optimize the constellation's peak-to-average power ratio (PAPR) for the in-phase and quadrature components, which increased the achievable transmitter SNR and partially reduced fiber
nonlinear interference\cite{mecozzi,Eric_ECOC}. Finally, the amplitude of the four outer constellation points were increased to form a distinct ring, enabling their use as markers for blind carrier phase estimation DSP.
The AWGN channel performance at each optimization stage, predicted by simulations, is shown in Fig.~\ref{fig:gap2cap} (constellations are shown inset). 

When viewing the constellation diagrams for GMI-optimized geometrically shaped modulation formats, it is important to note that, counter-intuitively, multiple constellation points can be coincident at certain coordinate values. 
This is because the gradient descent algorithm converged to a constellation where 8 discrete bit label pairs have similar IQ coordinates. As the trivial bit labelling is retained, the GMI-optimised 64-QAM constellations appears to have 56 points.

\section{Experimental configuration} 
%
%
%
%
\begin{figure*}[t]
   \begin{minipage}[b]{0.4\textwidth}
   \begin{tikzpicture}[
            const/.style={inner sep=0pt,anchor=center,draw,line width=1.3pt,rounded corners=8pt},
            ]
        	\begin{axis}[
            width=75mm,
            height=60mm,
            xlabel={SNR [dB]},
            ylabel={Gap to capacity [bit/4D symbol]},
            y label style={yshift=-0.1cm},
            xmin=0,xmax=20,
            ymin=0,ymax=1,
            legend cell align=left,
            legend style={font=\footnotesize},
            legend style={font=\footnotesize, at={(axis cs:0,1)}, anchor=north west},
            grid=major,
            every axis plot/.append style={line width=1.3pt},
            ]
        	
            \addplot[Set2-A] table[x=snr,y=square] {data/gap2cap.txt};
            \addlegendentry{Square 64QAM}

            \addplot[Set2-D] table[x=snr,y=awgn] {data/gap2cap.txt};
            \addlegendentry{AWGN tailored 64QAM}
            
            \addplot[Set2-B] table[x=snr,y=papr] {data/gap2cap.txt};
            \addlegendentry{PAPR tailored 64QAM}
            
			\addplot[Set1-A] table[x=snr,y=shaped] {data/gap2cap.txt};
            \addlegendentry{System tailored 64QAM}
            
            \draw[->,line width=1.3pt] (axis cs:11,0.57722) -- (axis cs:11,0.32465) node[midway, anchor=east] {\footnotesize 0.25 bit};
            
            \node[const,inner sep=-0.15cm,Set2-A] at (axis cs:2.0,0.12) {\includegraphics[width=1.3cm]{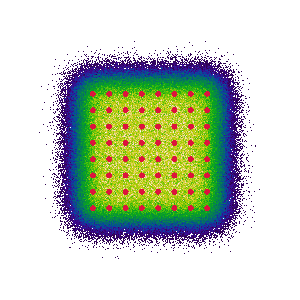}}; 
            \node[const,Set2-D] at (axis cs:7.0,0.12) {\includegraphics[width=1.cm]{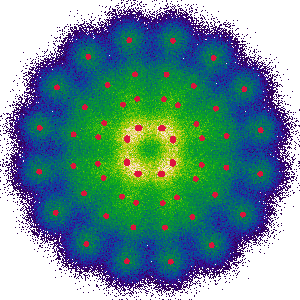}};
            \node[const,Set2-B] at (axis cs:12.0,0.12) {\includegraphics[width=1.cm]{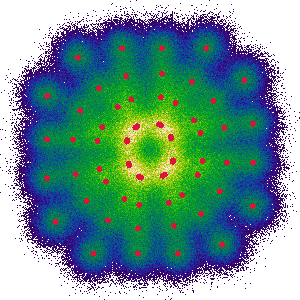}};
            \node[const,Set1-A] at (axis cs:17.0,0.19) {\includegraphics[width=1.6cm]{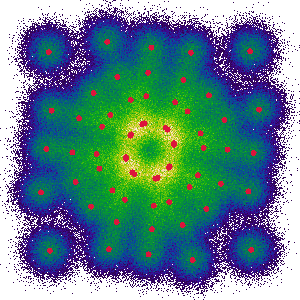}};
        	\end{axis}
        \end{tikzpicture}
        \vspace*{-2\baselineskip}
        \caption{Modulation formats optimized by received SNR.}
    \label{fig:gap2cap}
   \end{minipage}
   \hfill
   \begin{minipage}[b]{0.53\textwidth}
        \includegraphics[width=85mm]{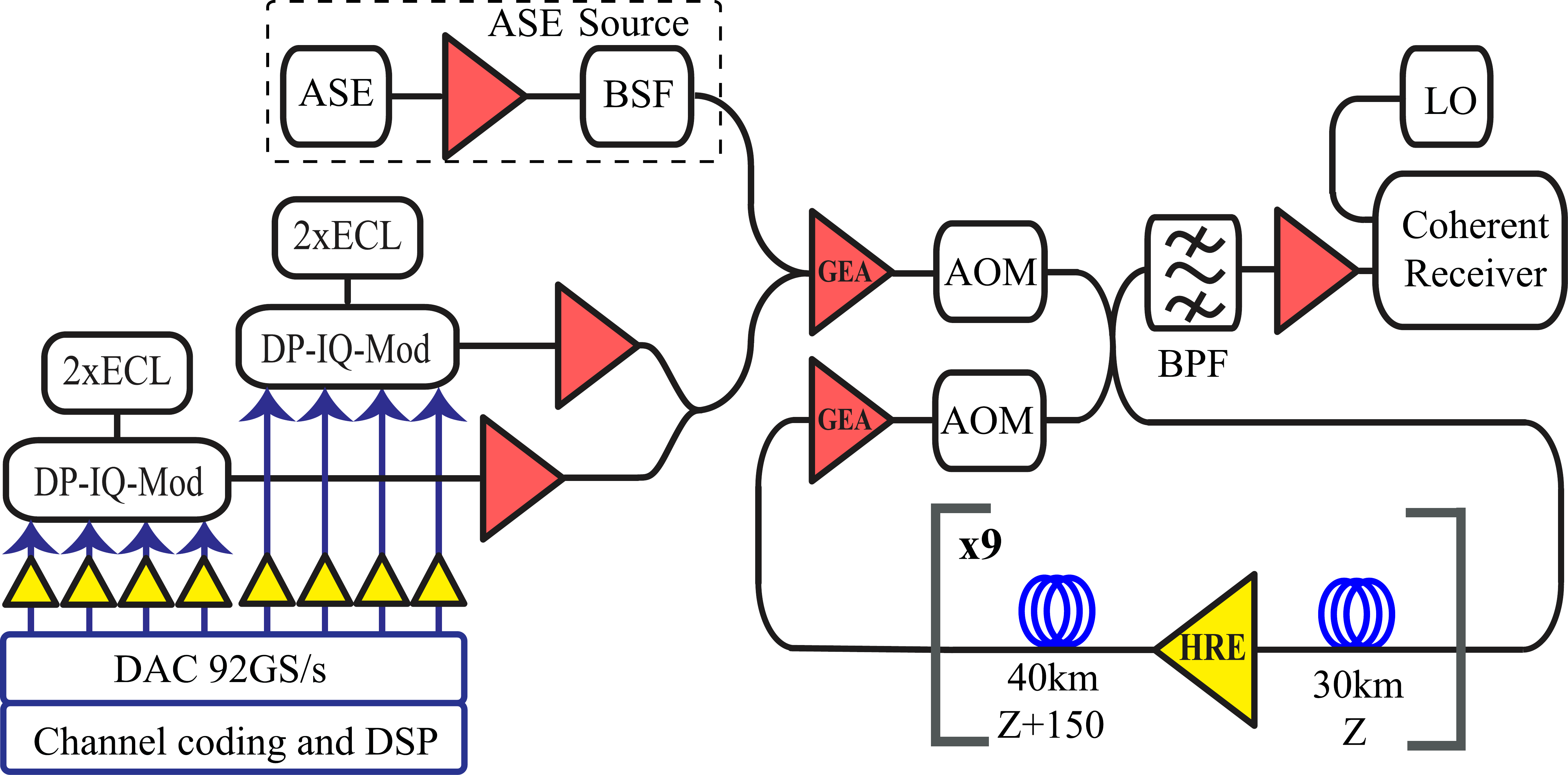}
    \caption{Experimental configuration.}
    \label{fig:setup}
    \end{minipage}
   
\vspace*{-\baselineskip}
\end{figure*}
The experimental configuration used for this work is shown in Fig.~\ref{fig:setup}. Four external cavity lasers (ECL) of 100~kHz linewidth were used to generate optical carriers spaced at 35.5~GHz. The carriers were modulated with the optimized geometrically-shaped 64QAM at 35~GBd by two independent dual-polarization IQ optical modulators, driven by four 92~GS/s digital-to-analogue converters (DACs) to generate four odd/even channels.
Digital root-raised cosine (RRC) filters with a 1\% roll-off were used to spectrally shape the signals. The transmitter boosting amplifiers were 97~nm bandwidth discrete Raman amplifiers with 12.5~dB gain. The channel carriers were tuned across the range 1525.56~nm  to 1614.54~nm, allowing the measurement of 306 channels, covering a total continuous spectrum of 89~nm. Spectrally shaped ASE noise of 97~nm bandwidth was used to emulate co-propagating channels over the entire transmitted bandwidth. A gain equalizing amplifier (GEA) with a continuous 91.04 nm bandwidth was used to amplify and spectrally shape the combined ASE and modulated channels to a -2~dB spectral tilt. 

The recirculating loop comprised 9 spans, each with 70 km of single-mode fiber and an HRE. Each span comprised two fiber types: a 40 km span of Sumitomo Z+150 fiber with an average attenuation of 0.148 dB/km and an effective core area of 149 $\mu m^2$ to limit nonlinear effects, followed by 30 km of Sumitomo Z fiber with an average attenuation of 0.16 dB/km and an effective core area of 81 $\mu m^2$ to provide sufficient Raman gain to achieve 0~dB gain across the spectrum.

After transmission over 6300 km, a bandpass filter was used to filter out channel 3 in the 4-channel group. Coherent detection was carried out using a phase- and polarization-diverse coherent receiver incorporating 70~GHz bandwidth photodetectors, and the signal was digitized using a four-channel real-time oscilloscope with a 63~GHz bandwidth, sampling at 160~GSa/s. 
The DSP block included matched filtering, chromatic dispersion compensation, a 19-tap blind radially-directive adaptive equalizer, frequency offset compensation and Viterbi $\&$ Viterbi carrier phase estimation \cite{VV1983}. For nonlinearity compensation, single channel digital back-propagation (DBP) was applied using 4~steps-per-span. After demapping, the log-likelihood ratios were passed to an off-the-shelf low-density parity check code from the DVB-S2X standard\cite{dvbs2x}. An outer BCH code was assumed with a 0.5\% overhead and a $3\times10^{-4}$ BER threshold.

\section{Results and Analysis}
Fig.~\ref{fig:throughput} shows the information rate of all 306$\times{}$35~GBd channels. The average received SNR before DBP was 11.09~dB with 7.23~bit/symbol. After DBP, the GMI was calculated giving an average of 7.49~bit/symbol pre-FEC, while the average post-FEC rate across all the channels was 6.93~bit/symbol.
%
%
The higher gains in the L-band are accounted for by the lower noise figure from Raman amplification.
After FEC, the average per channel net bit rate of 242.6~Gb/s providing a record single mode fiber capacity of 74.38~Tbit/s. Fig.~\ref{fig:post-FEC} shows that post-SD-FEC the BER of all channels are below the threshold.
Losses in the transmitter arrangement limited the back-to-back SNR to 20 dB, approximately equivalent to the SNR of two loops.
Thus, with improved transmitter SNR, a greater distance or higher capacity would be achievable.

\begin{figure}[thb]

\begin{minipage}[b]{0.4\textwidth}

\begin{tikzpicture}
\begin{axis}[
  every axis plot/.append style={line width=1.3pt,mark=*,only marks, mark size=1.3pt},
  grid=major,
  width=80mm,
  legend style={font=\footnotesize, , at={(axis cs:1520,4.5)}, anchor=west},
  xlabel={Wavelength [nm]},
  ylabel={Throughput [bit/symbol]},
  y label style={yshift=-0cm},
]


\addplot[Set1-A] table[x=wv,y=ngmi] {data/final_rate.txt};
\addlegendentry{GMI};

\addplot[Set1-B] table[x=wv,y=rfec] {data/final_rate.txt};
\addlegendentry{Post-FEC rate};

\end{axis}
\end{tikzpicture}
\vspace*{-2\baselineskip}
\caption{Throughput over 2 polarizations after 6300 km.}
\label{fig:throughput}

\end{minipage}
\hfill
\begin{minipage}[b]{0.5\textwidth}

\begin{tikzpicture}

\begin{semilogyaxis}[
  every axis plot/.append style={line width=1.3pt,mark=*,only marks, mark size=1.3pt},
  grid=major,
  width=80mm,
  legend cell align=left,
  legend style={font=\footnotesize, at={(axis cs:1530,0.01)}, anchor=west},
  xlabel={Wavelength [nm]},
  ylabel={Bit error rate [-]},
  y label style={yshift=-0.cm},
]

\addplot[Set1-A] table[x=wv,y=pre] {data/final_ber.txt};
\addlegendentry{Pre-FEC BER};

\addplot[Set1-B] table[x=wv,y=post] {data/final_ber.txt};
\addlegendentry{Post-FEC BER};

\addplot[Set1-B,mark=square,forget plot] table[x=wv,y expr=1e-6*\thisrow{postNaN}] {data/final_ber.txt};
\draw[->,line width=1.3pt,black] (axis cs:1570,3e-7) -- (axis cs:1570,9e-7) node[anchor=north west, font=\scriptsize] {No errors detected};

\draw[black,thick,dashed,line width=1pt] (axis cs:1510,3e-4) -- (axis cs:1630,3e-4);
\draw[] (1570,1e-3) -- (1570,1e-3, 1560) node[anchor = north, font=\scriptsize] {0.5\% HD-FEC OH};

\end{semilogyaxis}
\end{tikzpicture}
\vspace*{-2\baselineskip}
\caption{Pre-FEC and post-SD-FEC BER for all 306 channels.}
\label{fig:post-FEC}

\end{minipage}

\end{figure}

\vspace*{-2\baselineskip}
\section{Conclusions}
This work demonstrated a record transmission throughput for single-mode fiber over transoceanic distance.
Compared to the conventional C+L EDFA solution, hybrid EDFA/Raman amplification enables an extra 3~dB  loss per span and thus, fewer repeaters are required. The seamless transition from the C- to the L-band and lower noise figure enabled by these amplifiers, together with the optimized spectral tilt and system-tailored geometric constellation shaping, contributed to the overall capacity gain.

\vspace*{1\baselineskip}
\textit{Acknowledgements: This work was supported by UK EPSRC Program Grant TRANSNET EP/R035342/1 and the Royal Academy of Engineering under the Research Fellowships programme. The authors are grateful to Steve Desbruslais for fruitful discussions, Finisar and Sumitomo for their support in this investigation and Oclaro for the use of the high bandwidth, four-dimensional modulators.}


\end{document}